\documentclass{agujournal2019}
\usepackage{url} %this package should fix any errors with URLs in refs.
\usepackage{lineno}
\usepackage[inline]{trackchanges} %for better track changes. finalnew option will compile document with changes incorporated.
\usepackage{soul}
\usepackage{amsmath}
\usepackage{amssymb}
\usepackage{xcolor}
\usepackage{pdfpages}
\draftfalse

\journalname{Geophysical Research Letters}

\begin{document}

%%%%%%%%%%%%%%%%%%%%%%%%%%%%%%%%%%%%%%%%%%%%%%%
%  TITLE
%
% (A title should be specific, informative, and brief. Use
% abbreviations only if they are defined in the abstract. Titles that
% start with general keywords then specific terms are optimized in
% searches)
%
%%%%%%%%%%%%%%%%%%%%%%%%%%%%%%%%%%%%%%%%%%%%%%%

% Example: \title{This is a test title}

\title{Effect of submerged vegetation on water surface geometry and air–water momentum transfer}

%%%%%%%%%%%%%%%%%%%%%%%%%%%%%%%%%%%%%%%%%%%%%%%
%
%  AUTHORS AND AFFILIATIONS
%
%%%%%%%%%%%%%%%%%%%%%%%%%%%%%%%%%%%%%%%%%%%%%%%

% Authors are individuals who have significantly contributed to the
% research and preparation of the article. Group authors are allowed, if
% each author in the group is separately identified in an appendix.)

% List authors by first name or initial followed by last name and
% separated by commas. Use \affil{} to number affiliations, and
% \thanks{} for author notes.
% Additional author notes should be indicated with \thanks{} (for
% example, for current addresses).

% Example: \authors{A. B. Author\affil{1}\thanks{Current address, Antartica}, B. C. Author\affil{2,3}, and D. E.
% Author\affil{3,4}\thanks{Also funded by Monsanto.}}

\authors{G. Foggi Rota\affil{1}, A. Chiarini\affil{1,2}, and M. E. Rosti\affil{1}}

% \affiliation{1}{First Affiliation}
% \affiliation{2}{Second Affiliation}
% \affiliation{3}{Third Affiliation}
% \affiliation{4}{Fourth Affiliation}

\affiliation{1}{Complex Fluids and Flows Unit, Okinawa Institute of Science and Technology Graduate University, 1919-1 Tancha, Onna, Okinawa 904-0495, Japan.}
\affiliation{2}{Dipartimento di Scienze e Tecnologie Aerospaziali, Politecnico di Milano, via La Masa 34, 20156 Milano, Italy.}
%(repeat as many times as is necessary)

% Corresponding author mailing address and e-mail address:

% (include name and email addresses of the corresponding author.  More
% than one corresponding author is allowed in this LaTeX file and for
% publication; but only one corresponding author is allowed in our
% editorial system.)

% Example: \correspondingauthor{First and Last Name}{email@address.edu}

\correspondingauthor{M. E. Rosti}{marco.rosti@oist.jp}

%%%%%%%%%%%%%%%%%%%%%%%%%%%%%%%%%%%%%%%%%%%%%%%
% KEY POINTS
%%%%%%%%%%%%%%%%%%%%%%%%%%%%%%%%%%%%%%%%%%%%%%%
%  List up to three key points (at least one is required)
%  Key Points summarize the main points and conclusions of the article
%  Each must be 140 characters or fewer with no special characters or punctuation and must be complete sentences

% Example:
% \begin{keypoints}
% \item	List up to three key points (at least one is required)
% \item	Key Points summarize the main points and conclusions of the article
% \item	Each must be 140 characters or fewer with no special characters or punctuation and must be complete sentences
% \end{keypoints}

\begin{keypoints}
\item {First resolved simulation of turbulent air–water flow over vegetation, with quasi-realistic fluid properties.}
\item {Submerged vegetation smooths the water surface, damping steep deformations and regularizing streamwise wave fronts.}
\item {Despite altered interface shape, air–water momentum transfer remains unchanged, with implications for models.}
\end{keypoints}

%%%%%%%%%%%%%%%%%%%%%%%%%%%%%%%%%%%%%%%%%%%%%%%
%
%  ABSTRACT and PLAIN LANGUAGE SUMMARY
%
% A good Abstract will begin with a short description of the problem
% being addressed, briefly describe the new data or analyses, then
% briefly states the main conclusion(s) and how they are supported and
% uncertainties.

% The Plain Language Summary should be written for a broad audience,
% including journalists and the science-interested public, that will not have 
% a background in your field.
%
% A Plain Language Summary is required in GRL, JGR: Planets, JGR: Biogeosciences,
% JGR: Oceans, G-Cubed, Reviews of Geophysics, and JAMES.
% see http://sharingscience.agu.org/creating-plain-language-summary/)
%
%%%%%%%%%%%%%%%%%%%%%%%%%%%%%%%%%%%%%%%%%%%%%%%

%% \begin{abstract} starts the second page

\begin{abstract}

Understanding how submerged vegetation modifies the water surface is crucial for modeling momentum exchange between shallow waters and the atmosphere. In particular, quantifying its impact on the equivalent aerodynamic roughness of the water surface is essential for improved boundary-layer parameterization in oceanic and atmospheric models. In this Letter, we present fully resolved multiphase simulations of gravity-driven flow over a {fully} submerged vegetated bed, capturing the coupled dynamics of air, water, and individual plant stems, {under quasi-realistic conditions (the air/water viscosity ratio is real, while the density ratio is reduced tenfold)}. Our results show that vegetation {submerged for four times its height} regularizes the water surface suppressing strong deformations and homogenizing streamwise-propagating wave fronts along the transversal direction. Despite these alterations, the equivalent roughness perceived by the overlying air flow remains unchanged. These findings clarify vegetation–surface interactions and provide quantitative insights for nature-based wave mitigation strategies and atmospheric boundary-layer modeling.

\end{abstract}

\section*{Plain Language Summary}

The interaction between water and air at the surface of oceans, lakes, and rivers controls how gases like carbon dioxide and water vapor, as well as energy, are exchanged between the Earth’s surface and the atmosphere. In coastal regions, where aquatic plants such as seagrass are common, these interactions become more complex. Scientists and engineers need accurate ways to describe how submerged vegetation affects the roughness of the water surface and the motion of air above it to improve climate and weather models. In this study, {we use high-resolution computer simulations to investigate water flow over submerged vegetation, capturing the motion of water around individual plant stems and of the air above. We compare this case to an equivalent setup without vegetation. We find} that while the water surface appears smoother and more regular along the direction transversal to the flow when vegetation is present, it generates essentially the same drag on the airflow. These findings support the use of submerged vegetation in coastal protection strategies designed to reduce wave impact{, and they} help refine atmosphere–ocean models.

%%%%%%%%%%%%%%%%%%%%%%%%%%%%%%%%%%%%%%%%%%%%%%%
%
%  BODY TEXT
%
%%%%%%%%%%%%%%%%%%%%%%%%%%%%%%%%%%%%%%%%%%%%%%%

%%% Suggested section heads:
% \section{Introduction}
%
% The main text should start with an introduction. Except for short
% manuscripts (such as comments and replies), the text should be divided
% into sections, each with its own heading.

% Headings should be sentence fragments and do not begin with a
% lowercase letter or number. Examples of good headings are:

% \section{Materials and Methods}
% Here is text on Materials and Methods.
%
% \subsection{A descriptive heading about methods}
% More about Methods.
%
% \section{Data} (Or section title might be a descriptive heading about data)
%
% \section{Results} (Or section title might be a descriptive heading about the
% results)
%
% \section{Conclusions}

\section{Introduction}

Fluvial and coastal flows are often characterized by shallow layers of turbulent water moving over heterogeneous beds—ranging from smooth sand to rocky substrates, to complex surfaces covered with aquatic vegetation. Vegetated beds, or aquatic \textit{canopies}, play a critical role in limiting bed erosion \cite{zhao-nepf-2021}, modulating flow dynamics \cite{ghisalberti-nepf-2004}, and sustaining ecologically vital habitats \cite{duarte-etal-2020}. Their hydrodynamic impact has been extensively studied over the past decades \cite{nepf-2012-1}.

A fundamental distinction exists between emergent and fully submerged canopies. Emergent vegetation, such as mangrove forests, interacts with both the water column and the air–water interface, influencing wave propagation and surface morphology \cite{mei-etal-2011, mossa-depadova-2025}. Owing to their complexity, these systems are often studied through laboratory and field experiments \cite{liu-etal-2008, wu-cox-2015, vanRooijen-etal-2018, zhu-chen-2019}. In contrast, fully submerged canopies interact solely with the water column. Dense vegetation slows near-bed flow and generates a turbulent mixing layer at the canopy tip \cite{nepf-vivoni-2000, ghisalberti-nepf-2002}, where shear-driven turbulence replaces that generated by bed friction.

Much of our current understanding of submerged canopy flows derives from laboratory experiments \cite{poggi-etal-2004, ghisalberti-nepf-2006, tang-etal-2025} and fully resolved numerical simulations in open-channel configurations \cite{monti-omidyeganeh-pinelli-2019, lohrer-frohlich-2023, kim-yu-kaplan-2024, foggirota-etal-2024-2}. In both approaches, the air-water interface is typically neglected or idealized as a flat, frictionless boundary. Consequently, models have largely focused on vegetation-induced drag \cite{etminan-lowe-ghisalberti-2017} and subsurface mixing processes \cite{marjoribanks-etal-2017}, leaving the influence of vegetation on the air-water interface largely unexplored.

Yet the air-water interface remains dynamically coupled to the turbulence beneath. The extent to which submerged vegetation alters interface morphology and modulates momentum exchange with the atmosphere remains poorly quantified. This gap is particularly significant for atmospheric modeling, where surface roughness defines the lower boundary condition for turbulent momentum flux in the atmospheric boundary layer \cite{toba-1972, hwang-2022}. Understanding whether---and how---submerged vegetation modifies this roughness is therefore essential for accurately representing air-sea interactions in large-scale geophysical simulations \cite{drennan-taylor-yelland-2005, yang-meneveau-shen-2013}

\begin{figure}[htb]
\centering
\includegraphics[width=\textwidth]{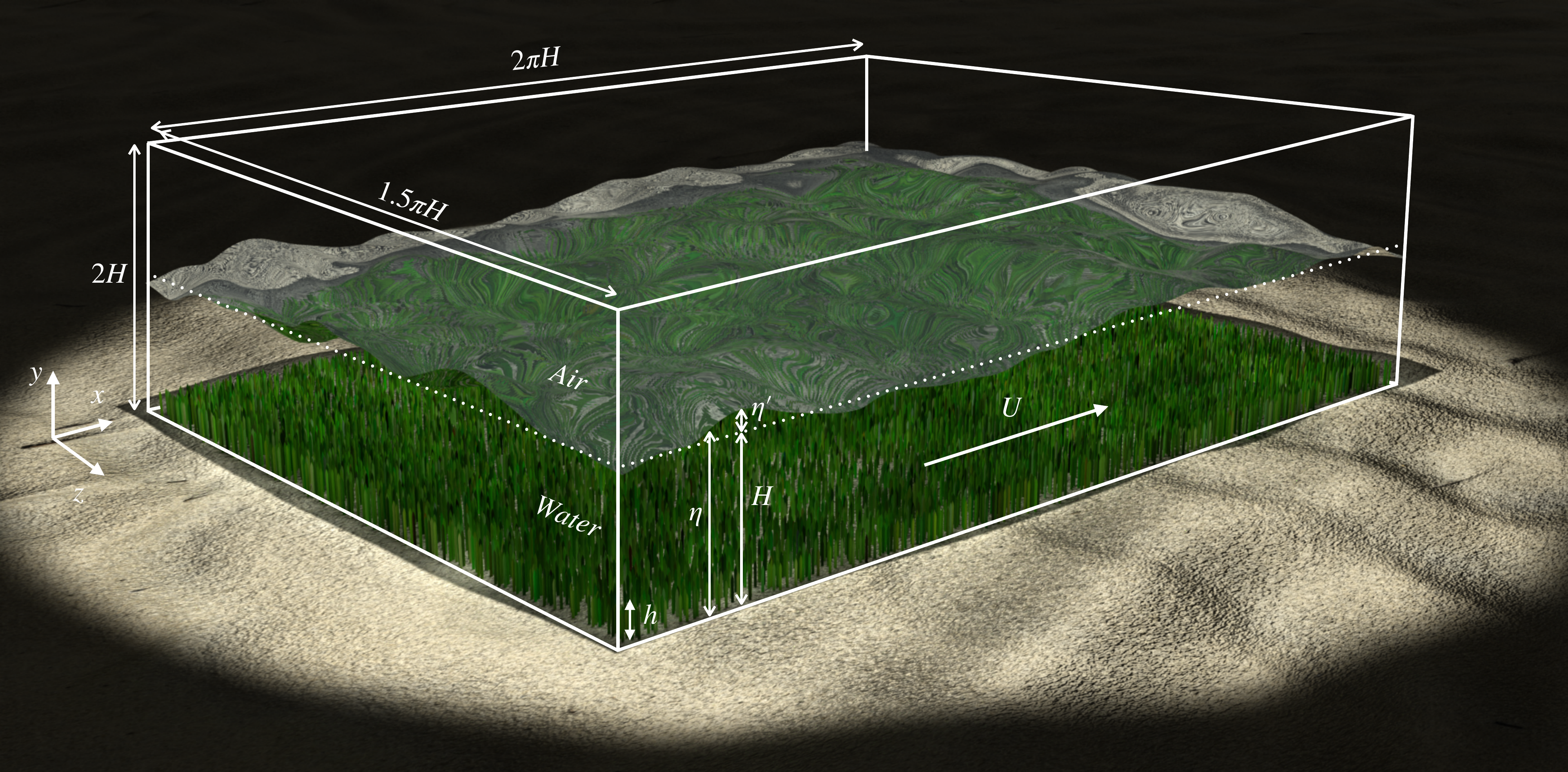}
\caption{Simulation setup, showing the individually resolved stems of the submerged canopy and the turbulent air–water interface above.
Relevant variables and parameters are sketched over an instantaneous realization of the system. Simulations capture the coupled dynamics of air and water, influenced by vegetation, to quantify how submerged canopies modify interface deformation.} 
\label{fig:cover}
\end{figure}

In this study, we investigate the influence of submerged vegetation on the water surface {in quasi-realistic conditions} using fully resolved numerical simulations that simultaneously capture air and water flows, a deformable air-water interface, and the detailed geometry of individual plant stems (Figure~\ref{fig:cover}). By comparing vegetated and smooth-bed cases under identical forcing, we show that vegetation significantly enhances surface deformation. Crucially, however, the overlying airflow remains largely unaffected. We further quantify the effective aerodynamic roughness length of the water surface as perceived by the atmospheric flow. These results have direct implications for understanding air-water coupling in vegetated systems, informing nature-based coastal protection strategies \cite{kobayashi-etal-1993, zhang-nepf-2021, zhu-etal-2022, foggirota-chiarini-rosti-2025}, and improving atmospheric boundary layer modeling.

\section{Numerical methods}

We perform simulations in a three-dimensional domain of length $2\pi H$, height $2H$, and width $1.5\pi H$ along the $x$, $y$, and $z$ axes, respectively (Figure~\ref{fig:cover}), {according to established numerical practice for canopy flows \cite{monti-olivieri-rosti-2023, sathe-giometto-2024}}. The domain is inclined at an angle $\Phi \approx 7^\circ$ relative to the vertical gravity vector $\mathbf{g}$. The lower half ($y \leq H$) is filled with water, the upper half ($y \geq H$) with air, placing the undeformed interface at $y = H$. No-slip, no-penetration conditions are imposed at the bottom wall, while the top boundary is free-slip. Periodic boundary conditions are applied in the horizontal directions. The gravitational incline drives a net flow along the positive $x$ direction in both fluid layers, {equivalently to the imposition of a pressure gradient along the streamwise direction. This configuration mimics the action of} a wind aligned with the underlying current.

The motion of the fluid is governed by the incompressible Navier–Stokes equations:
\begin{eqnarray}
&\nabla \cdot \mathbf{u} = 0,
\
&\displaystyle \frac{\partial \mathbf{u}}{\partial t} + \nabla \cdot (\mathbf{u} \mathbf{u})= \frac{1}{\rho_f}\left(- \nabla p + \nabla \cdot \boldsymbol{\tau}\right) + \sigma k \delta_s \mathbf{n} + \mathbf{f}_{\rm IBM} + \mathbf{g},
\label{eq:NS}
\end{eqnarray}
where $\mathbf{u}$ and $p$ are the velocity and pressure fields, $\rho _f$ is the local fluid density (equal to $\rho_w$ in water and $\rho_a$ in air), $\boldsymbol{\tau}$ is the viscous stress tensor, and further terms to the right represent body forces. Velocity is continuous across the air–water interface, while the stresses exhibit a jump due to surface tension. This effect is modeled as a volumetric force of the form $\sigma k \delta_s \mathbf{n}$ \cite{popinet-2018}, where $\sigma$ is the surface tension, $k$ the local curvature, $\mathbf{n}$ the normal vector, and $\delta_s$ a Dirac delta function concentrated at the interface. We also introduce a body force $\mathbf{f}_{\rm IBM}$ to account for the effect of vegetation elements.

The vegetated bed consists of $N_s = 15552$ cylindrical \textcolor{black}{rigid} stems of length $h = 0.25H$ and diameter $d \approx 2 \times 10^{-2} H$, arranged semi-randomly to avoid preferential flow channeling \cite{foggirota-etal-2024-2}. {To do so, we divide the bed into a grid of $n_x \times n_z = 144 \times 108$ rectangular tiles of area $\Delta S^2=(2\pi H/n_x)\times(1.5\pi H/n_z)$, and randomly place each stem within each tile sampling a uniform distribution as in \citeA{monti-olivieri-rosti-2023}. This tiling differs from the numerical grid and it is employed only to achieve the desired distribution of the stems, almost isotropic on the scale of the domain}. {The tips of the stems lay $3h$ under the water surface, at an ``intermediate" submergence level which is not deep nor shallow, where the vegetation effect on the surface is not trivially expectable}. Furthermore, the configuration considered produces a dense canopy \cite{monti-etal-2020} occupying approximately $4\%$ of the total simulation volume (air plus water), {which thus yields a marked modulation on the flow}.  Vegetation–fluid coupling is enforced via a no-slip condition using a force distribution $\mathbf{f}_{\rm IBM}$ computed through a Lagrangian immersed boundary method (IBM) \cite{huang-etal-2007, olivieri-etal-2020-2,foggirota-etal-2024}. Each stem is discretized with 126 uniformly distributed Lagrangian markers. Additional implementation details are provided in the Supplemental Material of \citeA{foggirota-etal-2024}.

Air and water phases are distinguished by a color function related to the volume-of-fluid (VOF) field $\phi(\mathbf{x},t)$ through cell averaging. This enables a monolithic formulation of equations~\eqref{eq:NS}, with fluid properties weighted by $\phi$ in each cell. The stress tensor is therefore computed as $\boldsymbol{\tau} = \mu_f (\nabla \mathbf{u} + \nabla \mathbf{u}^T)$, where $\mu_f(\mathbf{x},t)$ is the local dynamic viscosity. 

\begin{table}
\centering
\begin{tabular}{lcccccc}
\hline
 & $Re$ & $Fr$ & $We$ & $\mu_w / \mu_a$ & $\rho_w / \rho_a$ & $N_s$\\
\hline
Smooth bed     & 1000 & 3 & 6 & 55 & 80 & 0 \\
Vegetated bed  & 1000 & 3 & 6 & 55 & 80 & 15552 \\
\hline
\end{tabular}
\caption{Relevant flow parameters for the two simulated scenarios.}
\label{tab:params}
\end{table}

The system is characterized by several key dimensionless parameters, summarized in Table~\ref{tab:params}. Decomposing the gravity vector $\mathbf{g}$ into components parallel and normal to the bed, $g_x$ and $g_y$, we define a gravity-driven velocity scale $U = \sqrt{g_x H}$ {analogous to the friction velocity $u_\tau = \sqrt{\tau_w/\rho_f}$ in wall-bounded turbulence \cite{kim-moin-moser-1987}, where the wall shear stress $\tau_w$ balances a driving pressure gradient along the streamwise direction}. The inclination angle of the domain is then $\Phi = \arctan(g_x / g_y)$. The Reynolds number is defined as $Re = \rho_w U H / \mu_w$, where $\mu_w$ is the dynamic viscosity of water--- in its turn, analogous to the friction Reynolds number. We set $Re \approx 1000$ by adjusting $g_x$ to ensure fully turbulent conditions. The Froude number is fixed at $Fr = U / \sqrt{g_y H} \approx 3$ by appropriately selecting $g_y$. Surface tension is chosen to yield a Weber number $We = \rho_w U^2 H / \sigma \approx 1$, {as in \citeA{giamagas-etal-2023}}. The viscosity ratio is set to its physical value, $\mu_w / \mu_a = 55$, and a density ratio of $\rho_w / \rho_a = 80$ is imposed. While this density ratio is high for the practice of fully resolved simulations, where higher values lead to numerical instabilities, it is still tenfold away from reality. 
%Depleted air inertia might quantitatively alter the results discussed in the following, but we conjecture that they would not vary qualitatively as we are already considering two fluids with significantly different inertia (approximately two order of magnitude).
\textcolor{black}{An increase in the density ratio would likely affect the results quantitatively, but we can expect the changes to be minor and with no qualitative or phenomenological modifications. In fact, studies on bubble dynamics \cite{jain-etal-2019} have shown that limiting behaviours emerge for density ratios exceeding $\mathcal{O}(100)$. Moreover, in interfacial flows it is typically the Atwood number \cite{rigon-etal-2021}, rather than the density ratio itself, that governs inertia contrast. The Atwood number is defined as $A = (\rho_w - \rho_a)/(\rho_w + \rho_a)$, which in our simulations equals $A = 0.975$, only 2\% smaller than the realistic $A = 0.997$. Saturation is commonly observed for $A \gtrsim 0.94$ in the study of interfacial instabilities \cite{burton-2011}. Furthermore, our numerical setup compares well with the experiments of \citeA{shimizu-etal-1992}, performed in a canopy fully submerged in water with a free air-water interface above. In validating our solver, we carried out simulations in the water domain with a flat free-slip boundary at the top (full details are provided in the Supporting Information). The good agreement between these simulations and the experimental results confirms that our computations are performed in a regime where further increasing the inertia contrast at the interface plays a limited role.}  \vspace{-4.5pt}

We solve equation~\eqref{eq:NS} using our in-house, well-validated solver \textit{Fujin} (\url{https://groups.oist.jp/cffu/code}). Spatial derivatives are computed using second-order central finite differences, and time integration is performed with a second-order Adams–Bashforth scheme. Incompressibility is enforced via a projection–correction method \cite{kim-moin-1985}, where the pressure Poisson equation is solved efficiently using the \textit{2decomp} domain decomposition library, coupled with an in-place spectral solver based on Fourier series \cite{dorr-1970}. The volume-of-fluid (VOF) method adopts the multidimensional tangent of hyperbola for interface capturing (MTHINC) formulation by \citeA{li-etal-2012}, which has been validated against analytical benchmarks \cite{rosti-devita-brandt-2019} and applied in several recent studies \cite{devita-etal-2021,hori-etal-2023,cannon-soligo-rosti-2024}.  \vspace{-4.5pt}

Flow variables are discretized on a staggered Cartesian grid with 1152 uniform points in the streamwise ($x$) direction and 864 in the spanwise ($z$) direction. In the wall-normal ($y$) direction, 480 non-uniform grid points are used, to resolve the bottom-wall boundary layer and to maintain nearly isotropic cells near the air-water interface. This resolution is sufficient to accurately capture both smooth and vegetated bed cases, as demonstrated in prior studies \cite{monti-olivieri-rosti-2023, foggirota-etal-2024-2}. We have also verified that doubling the resolution does not significantly {alter the results}. \textcolor{black}{Each of the two simulations presented in this study was performed on 7,200 Fujitsu A64FX CPUs of the supercomputer Fugaku at RIKEN (\url{https://www.r-ccs.riken.jp/en/fugaku/}) for approximately three months: the first two months required to reach a fully developed turbulent state after a transient of $24 \sqrt{H/g}$, and the final month spent to collect sufficient statistics over $12\sqrt{H/g}$ time units, every $0.06\sqrt{H/g}$ for the fluid and every $0.6\sqrt{H/g}$ for the interface. Integration was performed with a time-step of $1.2\cdot10^{-5} \sqrt{H/g}$, and a total of 25,000,000 core-hours were employed for this project.}

\section{Results}
\subsection{Interface morphology}

\begin{figure}
\centering
\includegraphics[width=1\textwidth]{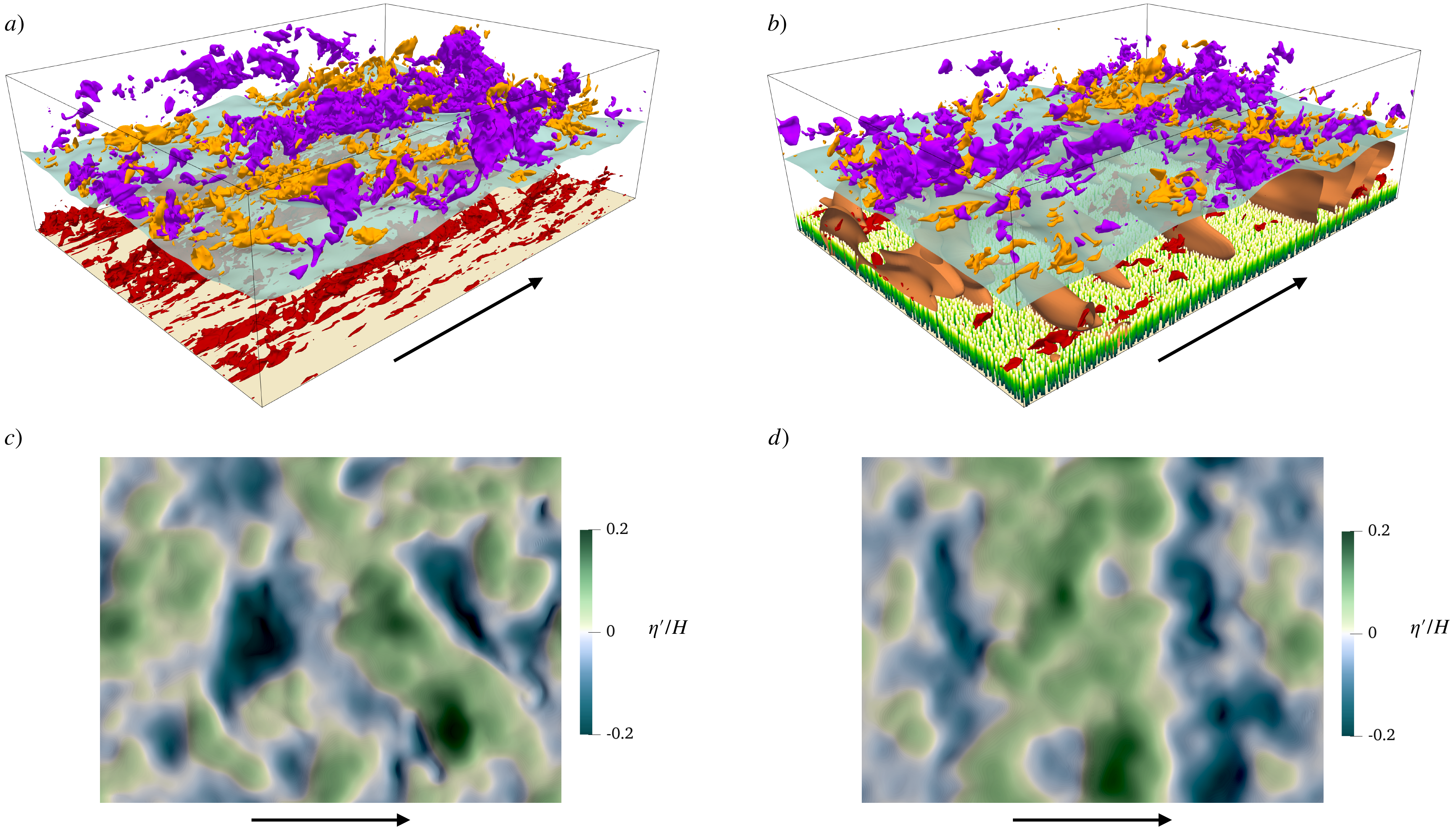}
\caption{Visualizations of the turbulent structures (panels~a,b) and interface deformation (panels~c,d) above the smooth and vegetated bed in the left and right panels, respectively. In (a) and (b), red iso-surfaces denote regions of high streamwise velocity in the water, while in the air orange iso-surfaces correspond to fast fluid moving towards the interface and purple iso-surfaces to slow fluid rising upward. {In (b) we also isolate Kelvin-Helmholts rollers in the water as brown iso-surfaces of the filtered pressure field (see main text for details).} In (c) and (d) we color interface fluctuations from $-0.2H$ to $0.2H$ seen from the top with a linear colormap ranging from blue to green, with white at null values. The interface appears qualitatively more rippled in the smooth bed case, while in the vegetated bed case we appreciate the spanwise regularization of streamwise--propagating wave fronts {over the underlying rollers}.} 
\label{fig:interface}
\end{figure}

A fully turbulent flow develops in our setup (see Supporting Information for further details) with velocity fluctuations in the water sustained either by the near-wall turbulent cycle over the smooth bed \cite{jimenez-pinelli-1999} or by the shear layer at the canopy tip \cite{nepf-2012-1}. This behavior is evident from the flow visualizations shown in panels~(a) and (b) of Figure~\ref{fig:interface}, where characteristic velocity streaks---central to the wall turbulence cycle---are observed near the smooth bed. Over the vegetated bed, by contrast, turbulence is dominated by Kelvin–Helmholtz rollers generated by the drag discontinuity at the canopy tip \cite{finnigan-2000, foggirota-etal-2024-2}. {In Figure~\ref{fig:interface} we isolate them as negative iso-surfaces of the pressure field smoothened with a box filter, following the same approach of our former study \cite{foggirota-etal-2025-1}}.
The turbulent fluctuations deform the air-water interface and propagate into the air layer, which exhibits sweep and ejection events {respectively defined as flow regions where $u'>0, v'<0$ or $u'<0, v'>0$, with $u'$/$v'$ the streamwise/vertical velocity fluctuations}. Fluctuation intensity peaks just above the interface and decays with height. The interface responds by bending and deforming under the action of turbulence, resulting in a complex and dynamic morphology that we aim to characterize.

{We compare flows over smooth and vegetated bed at fixed value of $Re_f$, aiming to isolate the effect of vegetation on the flow established over an otherwise unmodified incline.} Vegetation on the bed significantly alters the flow. 
Canopy drag reduces the mean velocity of the water, $U_{b,w}$, thereby decreasing the slip velocity at the interface, $U_{int}$ {(defined as the mean flow velocity at the average interface position $y=H$)}. As a result, the overlying air flow also slows down, with a corresponding reduction in its mean velocity, $U_{b,a}$. These values are summarized in Table~\ref{tab:RMS}.

\textcolor{black}{The vegetation considered in this study is rigid, while natural vegetation is typically flexible and therefore reconfigures \cite{alben-shelley-zhang-2002} and oscillates \cite{foggirota-etal-2024} under the action of the incoming turbulent flow. We thus forecast quantitative differences from our results in realistic environments, while the underlying physics should remain unchanged. Flexible stems, in facts, tend to assume a streamlined posture when they deflect (proportionally to their flexibility), thereby shielding the flow inside the canopy from the turbulent fluctuations above. The variable configuration of flexible stems diffuses the drag discontinuity generated at their tips, leading to weaker Kelvin–Helmholtz rollers compared to the rigid case and to an overall reduction in turbulence intensity in the outer flow \cite{monti-olivieri-rosti-2023, foggirota-etal-2024-2}. Thus, while stem flexibility significantly affects the flow inside the canopy, it leaves the outer turbulence attenuated but structurally similar. Since in our simulations the interaction between submerged vegetation and the water surface is primarily mediated by outer-flow turbulence, we choose to focus on the conditions where that turbulence is most intense and to consider rigid vegetation stems.}

\begin{table}
\centering
\begin{tabular}{lcccccccc}
\hline
 & $U_{b,w}/U$ & $U_{b,a}/U$ & $U_{int}/U$ & $\sqrt{\langle {\eta^\prime}^2 \rangle}$ & $\sqrt{\langle {\partial_x\eta^\prime}^2 \rangle}$ & $\sqrt{\langle {\partial_z\eta^\prime}^2 \rangle}$ & $\langle \mathcal{K} \rangle$ \\
\hline
Smooth bed      & 20.29 & 32.67 & 24.67 & 0.07 & 0.23 & 0.19 & 0.25 \\
Vegetated bed  & 4.95   & 18.29 & 10.24 & 0.07 & 0.20 & 0.14 & 0.31\\
\hline
\end{tabular}
\caption{Flow measurements and scalar indicators of the interface morphology, as introduced in the text. {Reported values are robust and converged, as they exhibit a variation less then $\sim5\%$ upon halving the time history used for their computation}.}
\label{tab:RMS}
\end{table}

Flow modulation \textcolor{black}{by the vegetated bed} leads to markedly different interface states.
Denoting the interface elevation from the bottom wall as $\eta(x,z;t)$, conservation of the two fluid phases requires that its mean value satisfies $\langle \eta \rangle = H$, where the operator $\langle \cdot \rangle$ denotes averaging over time and the homogeneous $x$ and $z$ directions. We therefore define the interface fluctuations as $\eta'(x,z;t) = \eta(x,z;t) - H$. This decomposition is illustrated in Figure~\ref{fig:cover}. Visualizations in panels~(c) and (d) of Figure~\ref{fig:interface} show that the interface appears more rippled over the smooth bed, while in the vegetated case, fluctuations are more homogeneous in the spanwise direction.
To quantify this, Table~\ref{tab:RMS} reports the root-mean-square (rms) of $\eta'$, which remains nearly unchanged between the two cases. However, the rms of its spatial gradients, $\partial_i \eta'$ with $i \in \{x, z\}$, reveals clear differences. Both gradients are attenuated over the vegetated bed, with the strongest reduction occurring in the spanwise ($z$) direction---consistent with the streamwise--propagating wavefronts observed in the visualizations.
{We further introduce a characteristic length-scale, $l_x$, defined as the first zero--crossing of the autocorrelation function $\mathcal{C}_{\eta,x}(\delta)=\langle \eta(x,z;t) \eta(x+\delta,z;t) \rangle / \langle {\eta^\prime}^2 \rangle$, and measure $l_x\approx1.2H$ in the vegetated bed case: a value $40\%$ larger than in the smooth bed case which confirms the interface regularisation by vegetation and might be interpreted as the typical wavelength of the streamwise--propagating wavefronts mentioned above.} Interestingly, although the canopy stems are {almost} isotropically distributed in the $x$–$z$ plane, the flow breaks this isotropy: submerged vegetation affects the interface differently along each direction. The spanwise regularization is likely linked to increased coherence of the underlying turbulent structures, particularly the formation of Kelvin--Helmholtz rollers \cite{finnigan-2000}.

\begin{figure}
\centering
\includegraphics[width=1\textwidth]{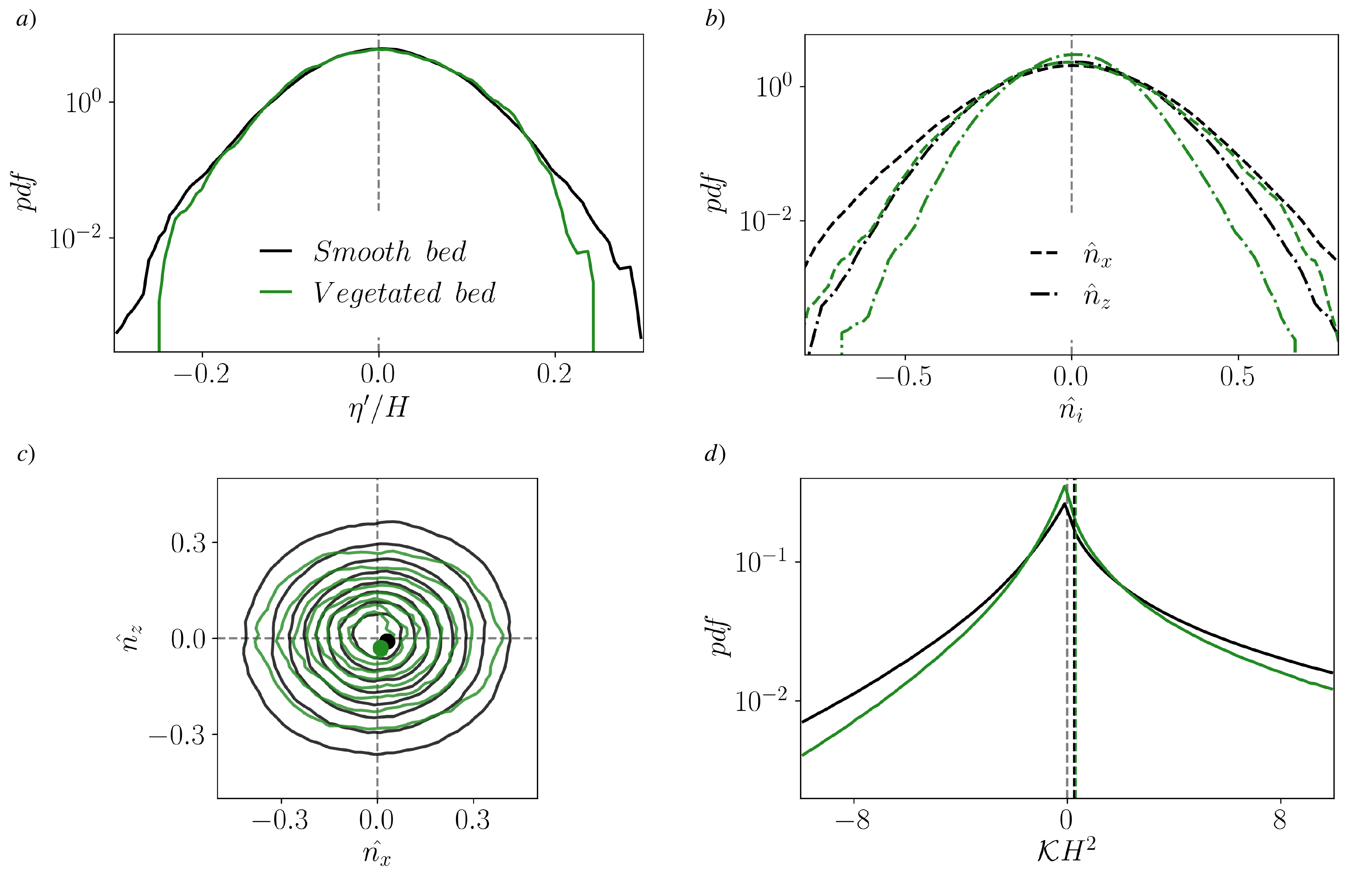}
\caption{Characterization of the interface deformation. Panel~(a) shows the probability density function of the interface elevation, while panel~(b) refers to the different components of the surface normal $\hat{n}$, {oriented upward}. In panel~(c) we present the joint probability density function of the normal components along the streamwise and spanwise directions, while in panel~(d) we report the distribution of the Gaussian curvature $\mathcal{K}$. Vegetation preferentially alters interface deformation along the transverse direction, reducing intense fluctuations.} 
\label{fig:pdfs}
\end{figure}

Further insight is provided by the probability density functions (PDFs) of the interface fluctuations. As shown in panel~(a) of Figure~\ref{fig:pdfs}, the distributions of $\eta'$ overlap at moderate amplitudes but diverge in the tails: the vegetated bed case consistently shows lower probabilities of large excursions, indicating a suppression of extreme interface deformations.

The surface gradients can be expressed through the surface-normal unit vector $\hat{\mathbf{n}} = \{\hat{n}_x, \hat{n}_y, \hat{n}_z\} = \{\partial_x \eta', 1, \partial_z \eta'\} / \sqrt{1 + \partial_x \eta'^2 + \partial_z \eta'^2}$, {pointing upward form the interface into the air}. The distributions of its components are shown in panel~(b) of Figure~\ref{fig:pdfs}. The effect of vegetation is most evident in the spanwise component $\hat{n}_z$, where the probability of steep interface slopes is markedly reduced. This trend is further confirmed by the joint PDF of $\hat{n}_x$ and $\hat{n}_z$ shown in panel~(c): while the smooth-bed case exhibits nearly circular isolines, the vegetated case appears compressed along the $\partial_z \eta'$ axis, reinforcing the anisotropic modulation of interface shape induced by submerged vegetation.

Additionally, we characterize the shape of the interface by computing its Gaussian curvature, $\mathcal{K}$, {expressed as the ratio of the determinants of the second and first fundamental forms, $\mathcal{K} = \det(\mathrm{II}) / \det(\mathrm{I})$ \cite{gauss-1828}}, and show it in panel~(d) of Figure~\ref{fig:pdfs}. At each point on the interface and at every time instant, $\mathcal{K}$ is defined as the product of the principal curvatures, $\kappa_1$ and $\kappa_2$, corresponding to the maximum and minimum curvatures of surface cross-sections taken in planes containing the local surface normal $\hat{\mathbf{n}}$. The distribution of $\mathcal{K}$ is skewed and exhibits a slightly positive mean (indicated by the colored vertical lines in panel~(d) and reported in Table~\ref{tab:RMS}), suggesting a predominance of dome-like deformations over saddle-like ones. However, the PDFs peak near zero curvature, consistent with the presence of propagating wavefronts characterized by one vanishing principal curvature ($\kappa_2 = 0$). As with other metrics, vegetation reduces the frequency of high-curvature events. 

We therefore conclude that the vegetated bed significantly influences the interface geometry, as evidenced by the visualizations in Figure~\ref{fig:interface}, globally suppressing intense deformations and regularizing streamwise-traveling wavefronts in the transverse direction. This finding is consistent with the preferential spanwise orientation of turbulent structures observed over vegetation canopies \cite{finnigan-2000}. 

\subsection{Momentum transfer to the overlaying air flow}

A conventional approach to quantify the dynamical influence of an irregular surface on the overlying turbulent flow---and thus the momentum transfer across the interface---is through the concept of \emph{equivalent roughness}, originally developed in the context of wall-bounded turbulence. The idea is to relate the effective roughness of a complex surface to that of the canonical experiments by \citeA{nikuradse-1933}, who measured flow resistance in pipes roughened with uniform spherical sand grains. Here, we extend this concept to assess whether the submerged canopy modifies the momentum exchange between water and air across the deformable interface.

The classical framework for describing momentum transfer near a boundary is based on the logarithmic law of the wall~\cite{pope-2000}, or simply log-law, which characterizes the mean velocity profile of a fully developed turbulent flow above a smooth wall:
\begin{equation}
\langle u^+ \rangle = \frac{1}{\kappa} \log \tilde{y}^+ + B,
\label{eq:logLaw}
\end{equation}
where $\langle u^+ \rangle=\langle u \rangle/u_\tau$ is the mean velocity expressed in viscous units, $u_\tau = \sqrt{\tau_w/\rho_f}$ is the friction velocity defined from the wall shear stress $\tau = \mu_f \text{d}\langle u \rangle/\text{d}|_y = 0$, and $\tilde{y}^+ = \tilde{y}/\delta_\nu$, with $\tilde{y}$ the distance from the wall and $\delta_\nu=\mu_f/(\rho_f u_\tau)$ the viscous length scale. The constants $\kappa \approx 0.41$ and $B \approx 5.2$ are empirically determined.
To account for surface roughness, the log-law can be reformulated using the concept of equivalent sand-grain roughness, denoted $k_s^+$
~\cite{jimenez-2004}, {to rescale the $\tilde{y}^+$ coordinate}. The quantity $k_s^+$ represents the radius of Nikuradse-type roughness elements that would produce the same mean velocity profile as the actual surface, thus serving as a dynamical measure of surface-induced momentum transfer. 
{In this case, the velocity profile becomes:}
\begin{equation}
{\langle u^+ \rangle = \frac{1}{\kappa} \log \left( \frac{\tilde{y}^+}{k_s^+} \right) + \tilde{B} \left( k_s^+ \right)}
\label{eq:logLaw2}
\end{equation}
{where $\tilde{B}$ is a roughness-dependent function obtained from Nikuradse's experiments. The limit $k_s^+ \rightarrow 0$ corresponds to a hydraulically smooth wall, with $\tilde{B} \rightarrow B + (1/\kappa) \log(k_s^+)$ and recovering Equation~\eqref{eq:logLaw}. Conversely, in the fully rough regime, $k_s^+ \rightarrow \infty$ and $\tilde{B}$ tends towards an universal constant $B_2$.}
When the roughness is irregular or non-uniform--as in the case of a deformable interface---$k_s^+$ no longer directly reflects geometric features but remains a meaningful dynamical parameter: it quantifies the effective momentum exchange between the surface and the overlying flow. Variants of this quantity are widely used in geophysical applications, where specifying the equivalent roughness of the water surface is critical for defining lower boundary conditions in atmospheric and oceanic models~\cite{drennan-taylor-yelland-2005, yang-meneveau-shen-2013}.
{
Assuming that the turbulent air flow over the water surface occurs in the fully rough regime, we adopt the modified form of the log-law considered by \citeA{yoshimura-omori-fujita-2024} to quantify momentum transfer across the air/water interface,
\begin{equation}
	\langle \Delta u^+ \rangle= \frac{1}{\kappa} \log{\frac{\tilde{y}+y_s}{y_s}}, \enspace y\ge H,
\label{eq:genLogLaw}
\end{equation}
reliant on the single parameter $y_s$.}
Here, $\tilde{y}=y-H$, $y_s$ plays the role of both a roughness length-scale and a vertical offset of the air flow origin over the surface{, and $\langle \Delta u^+ \rangle=\langle \Delta u \rangle / u_\tau$ with $\langle \Delta u \rangle = \langle u \rangle - U_{\text{int}}$ shifts} the mean air velocity profile by the interface advection speed \( U_{\text{int}} \) to reflect the moving boundary.
Unlike the case of a flat, rigid wall, the shear stress at the deformable interface is not solely due to viscous effects. The viscous shear stress, averaged across both water and air, is given by $\tau_\mu = \langle \mu_f \, \frac{\partial u}{\partial y} \rangle$. However, because velocity fluctuations at the interface are non-zero and surface tension contributes, two additional terms must be considered: the turbulent shear stress, $\tau_{\text{turb}} = -\langle \rho_f u' v' \rangle$, and the surface tension-induced stress \cite{rosti-devita-brandt-2019}, defined as $ \tau_\sigma = \int_0^{y} \langle \sigma k \delta_s n_x \rangle dy$. Thus, the total shear stress at any vertical location in the air is expressed as
\begin{equation}
\tau_{\text{tot}}(y) = \tau_\mu(y) + \tau_{\text{turb}}(y) + \tau_\sigma(y),
\label{eq:shearBalance}
\end{equation}
with no term vanishing at the interface $y=H$. We consequently generalize the friction velocity $u_\tau$ to be a local quantity varying with height, following the formalism of \citeA{monti-etal-2020}:
\begin{equation}
u_{\tau}(y) = \sqrt{\frac{\tau_{\text{tot}}(y)}{\rho_f \left( 2 - \frac{y}{H} \right)}}, \quad H \leq y \leq 2H.
\label{eq:frictionVel}
\end{equation}
This definition of $u_\tau $ is used to normalize the mean velocity profile in equation~\ref{eq:genLogLaw}. Note that the surface tension contribution $\tau_\sigma$ can be determined by enforcing closure of the shear balance in equation~\eqref{eq:shearBalance} once the driving force for the air flow is specified.

\begin{figure}
\centering
\includegraphics[width=\textwidth]{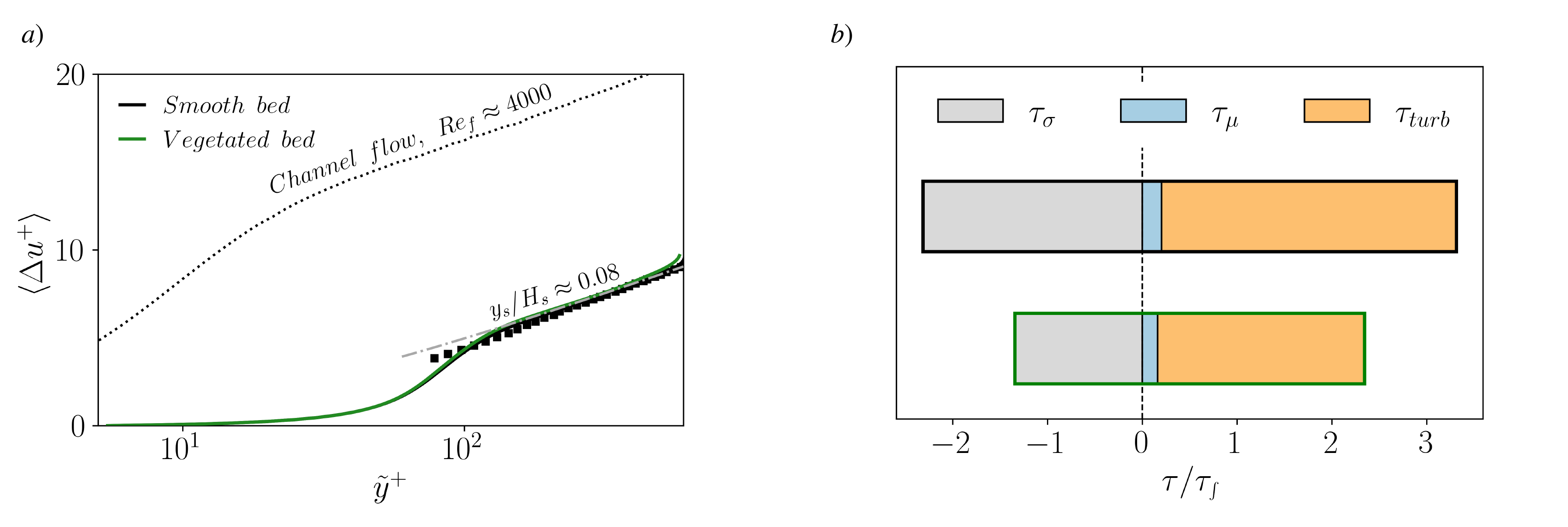}
\caption{Characterization of the surface roughness. In panel~(a) we report the mean profiles of the streamwise velocity in the air: the collapse of the two {curves} highlights the negligible effect of submerged vegetation on the air flow. Results from the simulations of \citeA{bernardini-etal-2014} over a flat rigid wall are reported as a dotted line, with the offset between those and our results quantifying the momentum deficit due to roughness. We report the fit with the log-law in equation \ref{eq:genLogLaw} as a gray dash-dotted line, and annotate the ratio between the equivalent roughness of the surface and the significant wave height above that. Our data are in good agreement with those of \citeA{sullivan-etal-2000}, reported in panel~(a) as black squares. In panel~(b) we show the integrated components of the shear stress balance (from equation~\ref{eq:shearBalance} in the text). } 
\label{fig:roughness}
\end{figure}

We are now positioned to quantify the roughness length scale $y_s$ in our setups. Under the scaling introduced above, the mean streamwise velocity profiles in the air collapse remarkably well {for the cases with the smooth and the vegetated bed}, as shown in panel~(a) of Figure~\ref{fig:roughness}. This indicates that, despite the significant morphological changes of the interface discussed in the previous section, the submerged vegetation considered here does not appreciably modify the turbulent air flow above. The momentum loss due to surface roughness is evident from the downward shift of the logarithmic region relative to the flat rigid wall case (i.e., channel flow), and is quantitatively characterized by $y_s$, which we measure as $y_s / H \approx 0.022$ regardless of vegetation presence. Conventionally, $y_s$ is expressed in terms of the \emph{significant wave height} $H_s$, defined as the average height of the highest one-third of waves in a given area, measured crest to trough. Following \citeA{goda-1970}, we estimate \textcolor{black}{$ H_s \approx 4 \sqrt{\langle \eta'^2 \rangle} \approx 0.28 H $} for both simulated scenarios, since $ \sqrt{\langle \eta'^2 \rangle} $ remains essentially unchanged (see table~\ref{tab:RMS}). There follows $ y_s / H_s \approx 0.08 $, which lies \textcolor{black}{close to the range reported by available field measuremens \cite{drennan-taylor-yelland-2005} and appears in remarkable agreement with the analytical prediction of \citeA{taylor-yelland-2000} (further details in the Supporting Information)}.  {Alternatively, in the formalism of equation~\eqref{eq:logLaw2} and assuming $\tilde{B}=11.8$, we find $k_s^+\approx 490$, confirming the former hypothesis of fully rough regime for which $k_s^+\gtrapprox 100$ \cite{kadivar-tormey-mcgranaghan-2021}}.

Our measurements in the logarithmic region align closely with those of \citeA{sullivan-etal-2000} (data were extracted from Figure~11 of their manuscript, for $c/u_* = 7.32$), shown as black squares in panel~(a) of Figure~\ref{fig:roughness}, {which they noted to lay at the threshold of the fully rough regime: a condition commonly encountered in geophysical flows \cite{kitaigorodskii-etal-1984}}. \citeA{sullivan-etal-2000} studied air flow exclusively, using a Couette setup where the velocity components at the water surface matched the irrotational orbital velocities of a two-dimensional deep-water gravity wave---a scenario markedly different from ours {and, notably, unaffected by the air/water density ratio}. Nevertheless, the close correspondence between their results and ours suggests that variations at the water surface {as well as an increase of the density ratio} have limited influence on the turbulent air flow within the logarithmic region.

The presence of the moving interface eliminates the conventional viscous region, where $\langle \Delta u^+ \rangle \approx y^+$ would typically hold. This peculiarity in our mean velocity profiles is better understood through the integral shear-stress balance, obtained by integrating equation~\eqref{eq:shearBalance} from the interface to the top of the domain, yielding $ \tau_\smallint = \int_H^{2H} \big( \tau_\mu(y) + \tau_{turb}(y) + \tau_\sigma(y) \big) dy$. {We analyze in panel~(b) of Figure~\ref{fig:roughness} the individual integral contributions normalized by $\tau_\smallint$, such that the histogram bars sum to unity.}
The viscous contribution $\tau_\mu$ plays a marginal role, consistent with the disappearance of the viscous sublayer, while the interfacial contribution $\tau_\sigma$ balances the turbulent fluctuations $\tau_{turb}$, justifying the rescaling introduced in equation~\eqref{eq:frictionVel}. Notably, $\tau_{turb}$---and consequently $\tau_\sigma$---is diminished in the vegetated case due to the reduced air flow over the interface, which moves at a lower velocity $U_{\text{int}}$, thereby lowering the effective Reynolds number of the air flow (see Supporting Information for further details). Nevertheless, this modulation is fully accounted for by our rescaling, leading to overlapping velocity profiles across the investigated scenarios for the same value of $y_s$.

\textcolor{black}{Turbulence generated at the canopy tip differs both qualitatively and quantitatively from that developing over the smooth bed, leading to distinct turbulent states in the water, which in turn imprint on the interface through the morphological modifications discussed in the previous section. However, introducing vegetation not only alters water-side turbulence, but also modifies the mean water flow, such that the air is exposed to a surface not only of different morphology, but also with different slip velocity. The framework we developed to rescale the mean flow profile of the air explicitly accounts for the change in slip velocity $U_{\text{int}}$, while recasting any residual dependence on interface morphology into the roughness length $y_s$. Remarkably, the finding that $y_s$ remains unchanged between the vegetated and smooth-bed cases demonstrates that the modulation of air-side turbulence, measured by $\tau_{turb}$, is entirely governed by the variation in slip, overshadowing the effects of the modest variation in the viscous shear stress $\tau_\mu$ shown in panel~b of Figure~\ref{fig:roughness}. In other words, interface slip sets the Reynolds number of the air flow over the interface. Upon introducing vegetation of the kind we considered, the interface slows down and turbulence is modulated according to the new Reynolds number, whose effect is accounted for by our rescaling. Compared to such effect, the role played by the modification of interface morphology, despite present, is negligible.
%While the submerged vegetation considered here does not appear to appreciably affect momentum transfer at the interface---leaving the roughness length $y_s$ unchanged---
We thus emphasize} the critical role of the interface velocity $U_{\text{int}}$ in the collapse of the mean velocity profiles $\langle \Delta u^+ \rangle$. Vegetation slows the water flow, which still exchanges approximately the same amount of streamwise momentum with the air as in the smooth bed case. As a result, the overlying air flow is also slower in the vegetated case. Given the similarity of the air-side velocity profiles, this reduction is fully captured by the velocity at the lower boundary of the air flow—the water surface. We report a reduction in $U_{\text{int}}$ of approximately 40\% between the smooth and vegetated bed scenarios.
From a modeling perspective, even if changes in surface roughness $y_s$ may be neglected, accurate knowledge of the interfacial velocity $U_{\text{int}}$ remains essential. While estimating $y_s$ requires detailed measurements of the air velocity profile in the logarithmic region---a challenging task---$ U_{\text{int}}$ can be readily obtained using simple experimental techniques, making it a more accessible and practical parameter for applications.

\section{Discussion}

We performed direct numerical simulations of two-phase turbulent flow over submerged vegetation, explicitly resolving both the air-water interface and all individual vegetation elements. These simulations approach realistic conditions and capture the essential dynamics of wave-current-vegetation interactions in aquatic environments.

We investigated the morphology of the water surface and the dynamics of the overlying air flow by comparing smooth-bed and vegetated-bed configurations under otherwise identical conditions. The key findings are as follows: 
(i) the presence of submerged vegetation significantly modifies the interface morphology, producing a geometrically smoother and more regular surface characterized by reduced curvature and surface gradients. In particular, interface fluctuations are noticeably regularized in the direction transverse to the mean flow; 
(ii) strikingly, despite these pronounced morphological alterations, the momentum transfer to the air remains essentially unaffected. This is evident when the mean velocity profile is appropriately rescaled using the local value of the total shear stress.
Thus, for the fully rough interface conditions considered here—where the air flow develops at a moderate distance from the vegetation canopy \cite{kitaigorodskii-etal-1984, sullivan-etal-2000}—the impact of submerged vegetation on the equivalent aerodynamic roughness may be neglected for atmospheric modeling purposes. In this regime, the only essential fitting parameter required to accurately {describe} the air flow is the velocity of the water surface.

These results advance our understanding of how aquatic vegetation influences water surface morphology, with implications for both fundamental research in environmental fluid mechanics and practical applications, such as the development of vegetation-based strategies for coastal protection \cite{kobayashi-etal-1993, zhang-nepf-2021, lorenzo-etal-2023}. Furthermore, by demonstrating the limited impact of submerged vegetation on the overlying air flow, this study provides valuable insights for atmospheric boundary-layer modeling.

Naturally, the scope of our work is limited to a specific configuration with a simple plant geometry at a fixed submergence level, {and a density ratio between water and air lower than the real one \textcolor{black}{(yet representative of a regime where the effect of the inertia contrast is saturated)}}. While the computational demand of simulations prevents us from a broader exploration of the parameter space, we anticipate that our conclusions may not hold when vegetation is significantly closer to the interface, potentially coming in contact with it. Future work should investigate this case, aiming to identify vegetation that optimizes interface modulation for wave attenuation purposes. In fact, the present canopy mimics a dense meadow of plants composed of single rigid stems---a somewhat idealized scenario---whereas alternative geometries, including irregular or patchy distributions \cite{winiarska-liberzon-vanhout-2024, park-nepf-2025, foggirota-etal-2025-1}, may offer improved or tunable performance for nature-based coastal defenses. \textcolor{black}{Additionally, natural vegetation is typically flexible, while our study only considers rigid plants. By doing so, we are able to clearly isolate the surface modulation induced by canopy-driven turbulence, while leaving open the question of any potential “resonant” interaction between surface waves and deflection waves within the canopy (i.e., \textit{monami} waves, \cite{ackerman-okubo-1993}), to be explored in upcoming studies.}

%%%%%%%%%%%%%%%%%%%%%%%%%%%%%%%%%%%%%%%%%%%%%%%
%
% DATA SECTION and ACKNOWLEDGMENTS
%
%%%%%%%%%%%%%%%%%%%%%%%%%%%%%%%%%%%%%%%%%%%%%%%

\section*{Open Research Section}
All data needed to evaluate the conclusions of this Letter are present in the main text and/or the supporting information.  
Processed data discussed in the text, used to generate figures and values reported in tables, are also made publicly available on Zenodo \cite{foggirota-chiarini-rosti-dataGRL-2025}: \url{https://doi.org/10.5281/zenodo.17588891}. Numerical simulations were carried out using a standard direct numerical simulation solver for the Navier–Stokes equations.  
Full details of the code implementation, validation, and related resources are available on the website of the Complex Fluids and Flows Unit at OIST (\url{www.oist.jp/research/research-units/cffu/fujin}).

\section*{Inclusion in Global Research Statement}
This research did not involve local communities, authorities, or field sites requiring permits or formal collaboration agreements. All data were generated through numerical simulations. The authors affirm their commitment to equitable research practices as outlined in the TRUST Code of Conduct.

\section*{Conflict of Interest Statement}
The authors declare there are no conflicts of interest for this Letter.

\acknowledgments
The research was supported by the Okinawa Institute of Science and Technology Graduate University (OIST) with subsidy funding to M.E.R. from the Cabinet Office, Government of Japan. M.E.R.~also acknowledges funding from the Japan Society for the Promotion of Science (JSPS), grants 24K00810 and 24K17210. The authors acknowledge the computer time provided by the Scientific Computing and Data Analysis section of the Core Facilities at OIST and the computational resources offered by the HPCI System Research Project with grants hp220402, hp240006, hp250035. \\

%%%%%%%%%%%%%%%%%%%%%%%%%%%%%%%%%%%%%%%%%%%%%%%
% REFERENCES and BIBLIOGRAPHY
%
% \bibliography{<name of your .bib file>} don't specify the file extension
% don't specify bibliographystyle
%
%%%%%%%%%%%%%%%%%%%%%%%%%%%%%%%%%%%%%%%%%%%%%%%

\bibliography{./Wallturb.bib}


\section*{Supplementary material}

**Effect of submerged vegetation on water surface geometry and air–water momentum transfer**

G. Foggi Rota[1], A. Chiarini[1,2], and M. E. Rosti[1]

[1]Complex Fluids and Flows Unit, Okinawa Institute of Science and Technology Graduate University, 1919-1 Tancha, Onna, Okinawa 904-0495, Japan.
[2]Current address, Dipartimento di Scienze e Tecnologie Aerospaziali, Politecnico di Milano, via La Masa 34, 20156 Milano, Italy.

**Contents of this file**

We characterize the turbulent flow attained in the water and in the air under the gravity forcing acting on the system. The underline{forcing Reynolds number} relative to the underline{water} phase, as defined in the main text, is imposed $Re \approx 1000$, while in the underline{air} $Re_a \approx 700$. Introducing the underline{bulk Reynolds number} $Re_{b,x} = \rho_x U_{b,x} H / \mu_x$ in the $x$ fluid phase, where $U_{b,x}$ is the mean velocity of the $x$ fluid phase along the streamwise direction, underline{over the bare bed} we measure $Re_{b,w} \approx 21000$ in the water and $Re_{b,a} \approx 23500$ in the air, while underline{over the vegetated bed} we have $Re_{b,w} \approx 5000$ and $Re_{b,a} \approx 13000$. In the following, figure S1 reports the mean velocity profiles and the turbulent fluctuations in the water scaled with $U_{b,w}$, while figure S2 shows the mean velocity profiles and the turbulent fluctuations in the air scaled with $U_{b,a}$. The fully turbulent nature of the air and water flows is also confirmed by the streamwise spectra of the turbulent kinetic energy $E_x(k_x)$, shown in figure S3. Further details are found in the captions of the figures.

We then underline{compare our results with available laboratory experiments, field measurements, and analytical predictions}. First, we confirm the capability of our numerical setup to correctly capture the flow dynamics within and above the canopy by performing simulations matching the parameters of the water-tank experiments by **Shimizu et al. (1992)**, who considered a rigid and dense canopy with height $h = 0.65H$ exposed to a turbulent flow at $Re_{b,w} = 7070$. underline{Our simulations accurately reproduce the mean flow features}, shown in the left panel of figure S4, as well as the Reynolds stresses (not shown). The same validation is also reported by *Monti et al. (2023)*. Next, we compare our reported value of the roughness length-scale $y_s$ with field measurements and analytical predictions available in the literature. **Drennan et al. (2005)** report field measurements in a parameter range close to that of our simulations, as shown in the right panel of figure S4, where $y_s/H_s$ is plotted against the wavelength $L_p$ at which the wave spectrum is most energetic, corresponding in our case to the computational domain length. underline{Our data point} lies at slightly higher $H_s/L_p$ due to the higher turbulence intensity of the simulated regime, yet it underline{compares well and shows remarkable agreement with the analytical prediction} of **Taylor and Yelland (2000)**.

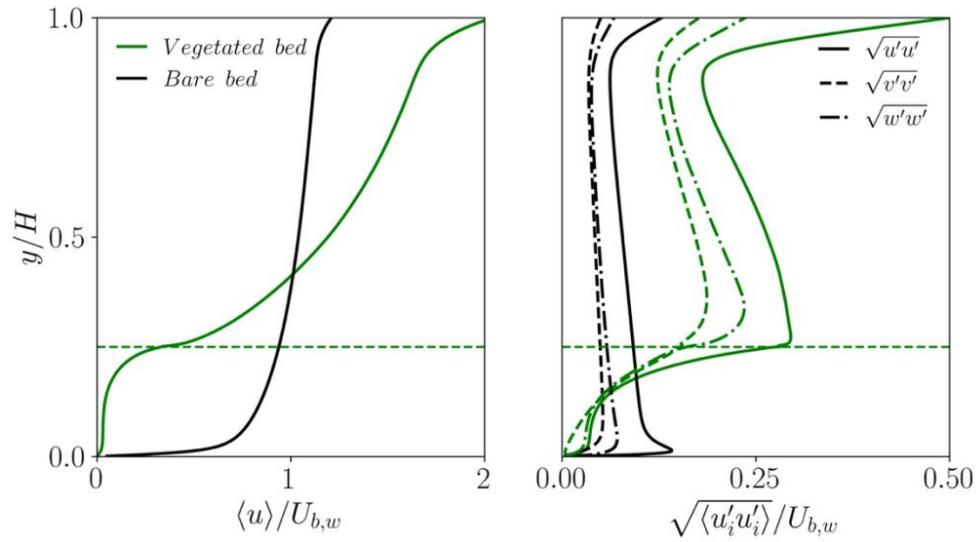

**Figure S1.** <u>Mean velocity profile (left) and turbulent fluctuations (right) in the water</u> flow below the interface, for the vegetated and bare bed case respectively. The double inflection in the mean velocity profile of the vegetated case is consistent with results for dense canopies *(H. M. Nepf, Annu. Rev. Fluid Mech. 2012)*, while the fluctuations confirm the fully turbulent nature of the flows.

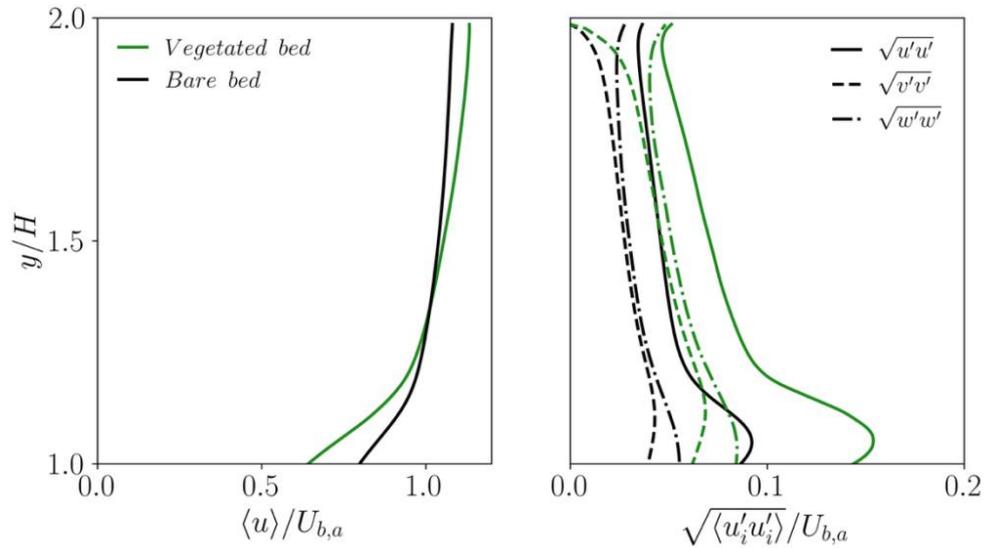

**Figure S2.** <u>Mean velocity profile (left) and turbulent fluctuations (right) in the air</u> flow above the interface, for the vegetated and bare bed case respectively. Note the steeper gradient $\partial_{\boldsymbol{y}}\langle\boldsymbol{u}\rangle$ of the mean velocity profile at the interface of the vegetated case compared to the bare bed one, yielding the increased shear mentioned in the main text.

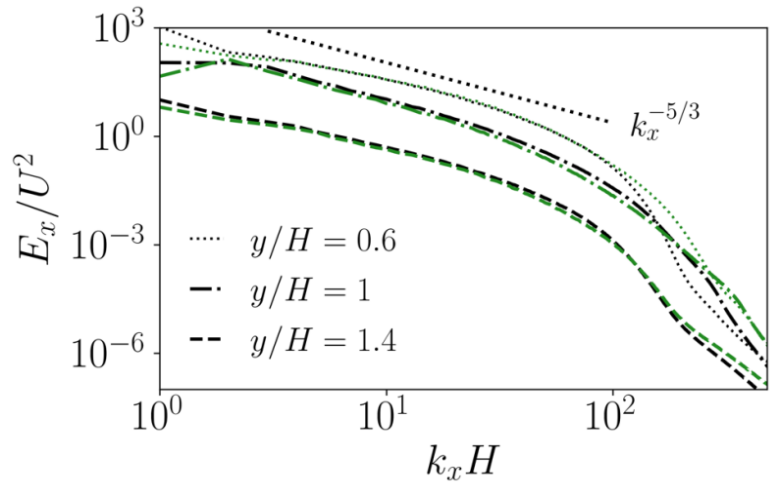

**Figure S3.** Turbulent kinetic energy spectra along the streamwise direction in the water ($y/H = 0.6$), at the interface ($y/H = 1$), and in the air ($y/H = 1.4$). The smooth bed and the vegetated bed cases are distinguished with black and green color, respectively. Spectra at $y/H = 0.6$ are shifted up of a factor 100 and spectra at $y/H = 1$ of a factor 10, to avoid overlapping in the plot. The fully turbulent nature of the flow is confirmed by the fact that, at all vertical positions, almost two decades of inertial range exhibiting $E_x \sim k_x^{-5/3}$ are attained.

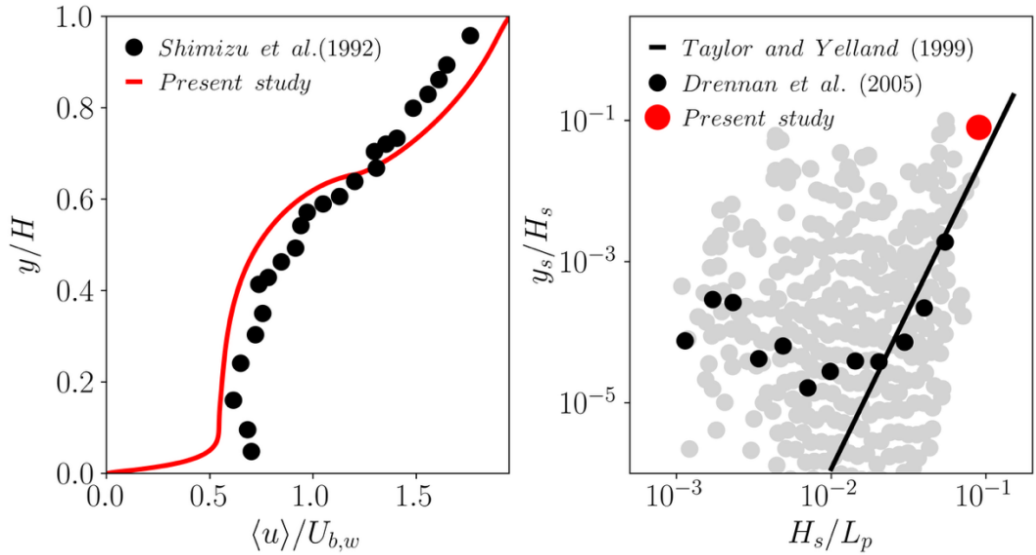

**Figure S4.** Validation against laboratory and field measurements. Mean streamwise velocity profile in and above a dense, rigid canopy (left), where black denotes experimental measurements *(Shimizu et al., 1992)* and red the outcome of a simulation with our code *(Monti et al., 2023)*, matching the experimental parameters. Dimensionless roughness $y_s/H_s$ versus significant steepness $H_s/L_p$ (right) from the present study, field measurements *(Drennan et al., 2005)*, and analytical prediction *(Taylor and Yelland, 2000)*. Gray dots denote measurements from *Drennan et al. (2005)*, while black dots correspond to their binned average.



\end{document}